# TMIC: App Inventor Extension for the Deployment of Image Classification Models Exported from Teachable Machine

**Fabiano Pereira de Oliveira**
Department of Informatics and Statistics, Federal University of Santa Catarina
Florianópolis/SC, Brazil
fabiano.pereira.oliveira@grad.ufsc.br

**Christiane Gresse von Wangenheim**
Department of Informatics and Statistics, Federal University of Santa Catarina
Florianópolis/SC, Brazil
c.wangenheim@ufsc.br

**Jean C. R. Hauck**
Department of Informatics and Statistics, Federal University of Santa Catarina
Florianópolis/SC, Brazil
jean.hauck@ufsc.br

**Summary**

TMIC is an App Inventor extension for the deployment of ML models for image classification developed with Google Teachable Machine in educational settings. Google Teachable Machine, is an intuitive visual tool that provides workflow-oriented support for the development of ML models for image classification. Aiming at the usage of models developed with Google Teachable Machine, the extension TMIC enables the deployment of the trained models exported as TensorFlow.js to Google Cloud as part of App Inventor, one of the most popular block-based programming environments for teaching computing in K-12. The extension was created with the App Inventor extension framework based on the extension PIC and is available under the BSD 3 license. It can be used for teaching ML in K-12, in introductory courses in higher education or by anyone interested in creating intelligent apps with image classification. The extension TMIC is being developed by the initiative Computação na Escola of the Department of Informatics and Statistics at the Federal University of Santa Catarina/Brazil as part of a research effort aiming at introducing AI education in K-12.

**Statement of need**

Nowadays, Machine Learning (ML) is present in our daily lives, e.g., as spam filters, recommendation mechanisms, chatbots, digital assistants, etc. Considering its inevitable impact it is essential that people understand Machine Learning not just as a consumer, but also as a creator of this type of innovation (Touretzky et al., 2019). Therefore, it is important to start teaching ML concepts in K-12 following a trend emerging during the last few years (Marques et al., 2020).
According to the *K-12 Guidelines for Artificial Intelligence* by AI4K12 (Touretsky et al., 2019), AI education encompasses 5 big ideas, including Machine Learning (ML). This includes an understanding of basic ML concepts as well as the application of these concepts, developing ML applications typically focusing on the task of image classification. Image classification is the process of taking an input (like a photo or videostream) and outputting a class (like "plastic garbage") or a probability that the input is a particular class ("there's a 90% probability that this image shows a plastic garbage"). Teaching the application of ML following a human-centric interactive ML process (Amershi et al., 2019, Gresse von Wangenheim and von Wangenheim, 2021) includes the development, from requirements analysis to exporting the developed model and its deployment (Figure 1).

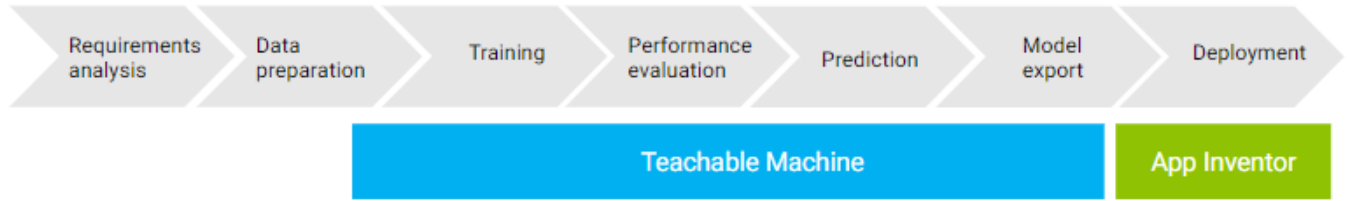

Figure 1. Human-centric interactive ML process

The development of ML models is typically taught in K-12 adopting visual tools, such as Google Teachable Machine. Google Teachable Machine (teachablemachine.withgoogle.com) is a free web-based GUI tool for creating custom machine learning classification models without specialized technical expertise (Carney et al., 2020). Google Teachable Machine uses TensorFlow.js, to train and run the users' models online. It runs within the browser entirely on the user's device, maintaining any input data locally, protecting data privacy. Google Teachable Machine provides an intuitive workflow-oriented interface supporting the upload of the dataset, model training and evaluation, as well as the prediction of the classes of new data and the export of the trained model (Figure 2).

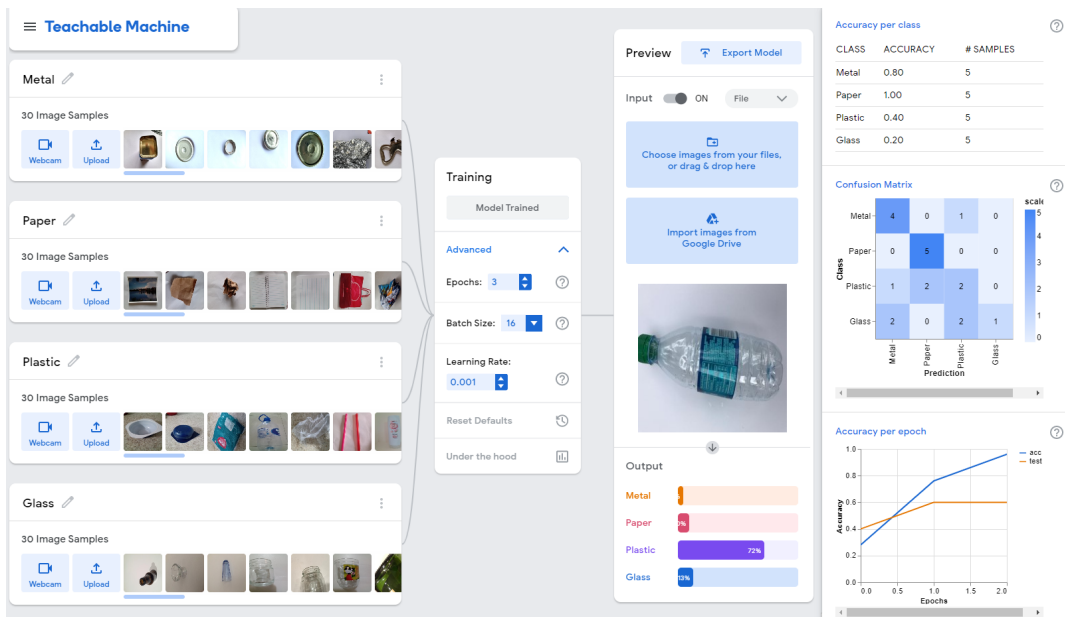

Figure 2. Example of ML model development with Google Teachable Machine

Furthermore, it is possible to download a trained model as a TensorFlow.js model and host it on Google Cloud, so it can be deployed easily into any website or app. In this case Google Teachable Machine generates a URL where the model is hosted for free and this link can be shared to use the created model. It can also be converted into a TensorFlow or TensorFlow Lite model and downloaded for local use (Figure 3).

Figure 3. Example of exporting trained model as TensorFlow.js

The exported Tensorflow.js is a zip file including:
- **Metadata**, a JSON file indicating the versions of the used libraries, metadata on the user and model name, as well as a list of the label names and the size of the images used to train the model
- **Model,** a JSON file specifying the model topology
- **Weights**, a BIN file specifying the weights of the trained model

The exported trained ML model can then be deployed in software systems, such as mobile applications. The deployment of the trained model is important to illustrate the usefulness of ML, not only teaching the development of ML models, but also the creating of "intelligent" solutions. Such a deployment as part of IA/ML education in K-12 is typically done within block-based programming environments such as App Inventor (Gresse von Wangenheim et al., 2021).

MIT App Inventor (appinventor.mit.edu) is a free web platform that allows users to create mobile applications. Users can design their own applications using drag and drop components and program its behavior using a blocks-based programming language. The App Inventor core already provides a comprehensive set of components, methods and commands for diverse kinds of functionality, including sensors, communication, data storage, etc. It is also possible to further extend App Inventor by providing more components (Patton et al., 2019).

Extensions can also be used to incorporate ML features into App Inventor using the App Inventor extension framework (http://ai2.appinventor.mit.edu/reference/other/extensions.html). An example of such an App Inventor extension for integrating custom-trained image classification models is the Personal Image Classifier (PIC) extension (Tang et al., 2019)(Tang, 2019). Support provided by PIC consists of a web application supporting the model development and of the extension with new components for running the trained model in App Inventor apps. However, certain shortcomings regarding the specific web application for the development of the model, such as a lack of evaluation support in V2.0, performance problems of the trained models, and a lack of flexibility allowing the deployment of ML models developed on other environments such as Teachable Machine, indicate a need for further extensions.

**TMIC Teachable Machine - Image Classifier Extension**

TMIC is an App Inventor extension for the deployment of image classification models developed in Google Teachable Machine. The extension is based on the PIC extension (Tang et al., 2019)(Tang, 2019), adapting the PIC extension in order to enable the import of TensorFlow.js models created with Google Teachable Machine and exported by uploading them on Google Cloud.

The TMIC extension includes the following properties:

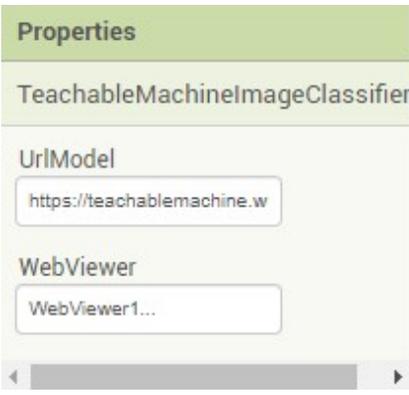

**URL_Model** is the property responsible for containing the URL of the model trained with the Google Teachable Machine that has been exported as Tensorflow.js on Google Cloud.

**WebViewer** is the property that allows the user to assign a Web Browser component so that the extension can be used. The Web Browser component is used in order to visualize the functionality of the extension.

The TMIC extension provides the following blocks:

| Blocks of the TMIC extension | Functionality |
| --- | --- |
| when TeachableMachineImageClassifier1 .ClassifierReady do | The ClassifierReady event block is executed when the extension finishes loading the ML model from the GTM cloud. |
| when TeachableMachineImageClassifier1 .GotClassification result do | The GotClassification event block is executed when the extension finishes classifying an image. This event occurs right after the execution of the ClassifyVideoData block, returning the list of predictions for each category in the model. |
| call TeachableMachineImageClassifier1 .ClassifyVideoData | The ClassifyVideoData block starts the classification of the image captured by the smartphone's rear-facing camera video stream, using the WebViewer component. When the classification is finished, the result is returned via the GotClassification event block. |
| 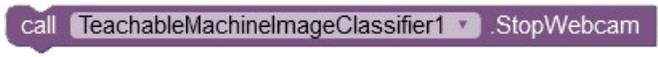 | The StopWebcam block stops the webcam when leaving the screen in which the image classification is done. |
| 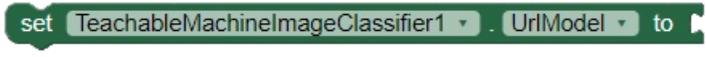 | The URL_Model adjustment block allows the user to adjust the ML model URL to another link of the exported GTM model in Google Cloud. |
| 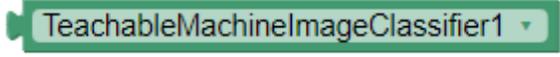 | The TeachableMachineImageClassifier get block returns a specific instance of the extension. |

The TMIC extension was developed using as a base the PIC extension code inside of the App Inventor framework for creating extensions. It is divided into backend and frontend, in which the first refers to the TeachableMachineImageClassifier.java file, which initializes the extension and its blocks.

Each extension block is defined in the TeachableMachineImageClassifier.java class as a method responsible for executing the action of this block.

The front end of the TMIC extension is made up of an HTML file along with four other Javascript files. The extension needs to be rendered in an HTML page to open the camera (asking the user for permission), and display it to the user. Javascript files are needed to load the model and perform the classification of the image when the user requests it. The teachable_machine_image_classifier.js file is mainly responsible for performing the task of communicating with the backend, receiving requests for loading the model, opening the cell phone camera and classifying the image. This file also notifies the backend when the extension is ready and returns the ranking results.

The other Javascript files refer to the GTM and TensorFlow.js frameworks, which work together to perform image classification on the mobile device. Its functions are called within the teachable_machine_image_classifier.js code and internally between the files. In total, the TMIC extension is made up of six files, one of the Java extension, one of the HTML and four Javascript extension files.

The main methods of the Java class (TeachableMachineImageClassifier.java) are those that will define the behavior of the blocks. Most of them need to communicate with the front-end, calling functions and passing parameters to the file teachable_machine_image_classifier.js, which will process the submitted request.

The teachable_machine_image_classifier.js Javascript file functions are responsible for receive requests from the TeachableMachineImageClassifier.java class, process them and return the result in cases where it is necessary to notify that the extension is ready for the usage or prediction results are ready.

The classifyVideoData() function is responsible for loading the Teachable Machine cloud hosted template from a user-defined URL in the property URL_Model. The classifyVideoData() function captures the URL of the two JSON files, one containing the model trained and the other with the model metadata, from the URL defined earlier in the preparation of the extension when it is initialized. After assigning the URL of the JSON files to two variables, they are passed as parameters to the function tmImage.load(modelURL, metadataURL), belonging to the Teachable Machine framework file teachablemachine-image.min.js, which will return the model prepared to be used in image classification.

After the model is prepared, the model.getTotalClasses() function is called, which returns the number of model classes for the maxPredictions variable. Finally, the function is called model.predict(webcamPredict.canvas), which calls the classifier passing as a parameter the image captured at the exact moment the classifyVideoData() function was executed, returning the prediction result in the prediction variable. Finally, after the prediction variable is already with the classification result, an array with the results is filled in and returned to the Java class, calling the method reportResult(String result), which will prepare the result and notify the event block GotClassification.

The TMIC extension is provided with the BSD 3 license (https://opensource.org/licenses/BSD-3-Clause), included in a LICENSE file. The license is available from the TMIC source code in the GitLab repository hosted at the Federal University of Santa Catarina. Also along with the code sources, a NOTICE file is available recognizing that the TMIC was developed by adjusting the PIC code.

Currently the TMIC extension supports only Tensorflow.js models exported to Google Cloud and only allows capturing images with the rear-facing camera of the smartphone. We are planning to improve the extension as part of future work.

**Usage example**

The extension can be used for teaching ML in K-12, in introductory courses in higher education or by anyone who wants to create "intelligent" apps for image classification. As with any App Inventor extension it can be imported into App Inventor and then used in order to run trained models as part of intelligent apps.
In order to support its usage the following material is available (currently in Brazilian Portuguese only):

- TMIC extension .aix
- Example app for the classification of recycling trash .aia (wireframe and final version)

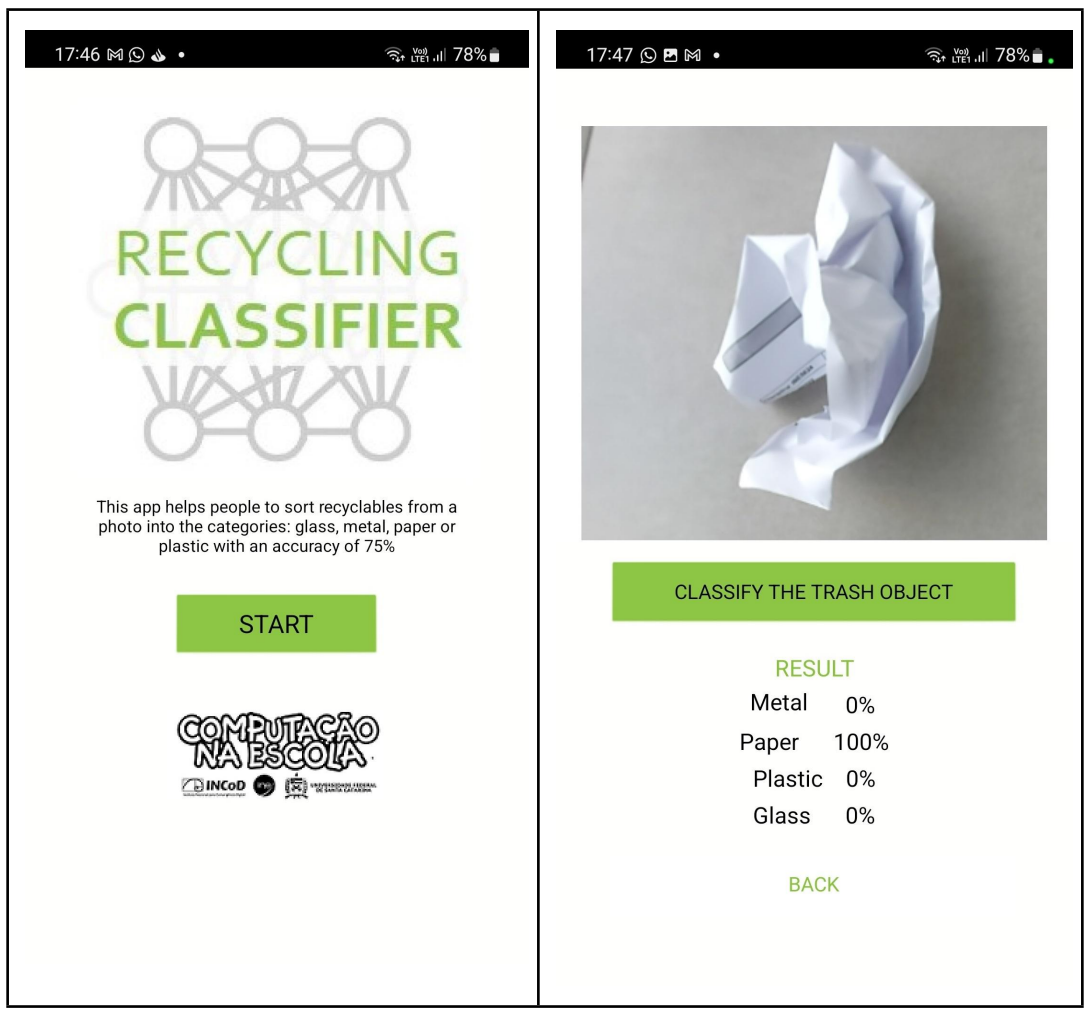

- Online tutorial explaining the use of the extension

The extension and material is available online: https://computacaonaescola.ufsc.br/en/tmic/

## Acknowledgments

This work is supported by CNPq (National Council for Scientific and Technological Development), a Brazilian government entity focused on scientific and technological development.